# The dark resonances in the case of the absorption of optical radiation in double tunneling-coupled quantum wells

A.N. Litvinov[1], K.A Barantsev[1], B.G. Matisov[1], G.A. Kazakov[1],

Yu.V. Rozhdestvensky[2]

[1]St. Petersburg State Polytechnical University, 195251 St. Petersburg, Russia

e-mail: andrey.litvinov@mail.ru

[2]State University of Information Technologies, Mechanics and Optics,

197101 St. Petersburg, Russia

**Abstract:** This work is devoted to the research of conditions of the dark resonance (coherent population trapping) of the interaction of the laser radiation with tunneling-coupled quantum wells. The phase sensitive dependence of dark resonances has been investigated. We obtained that destruction as well as restoration of the dark resonances of the coherent population trapping is possible in dependence on algebraic sum of the phases of exciting fields. It is shown that the phase variation of exciting fields influences on the absorption and dispersion of the probe field in a medium with quantum wells.

## 1. Introduction

Now the coherent population trapping (CPT) effect attract big interest of researchers which consist in arising of superposition quantum state (dark state) of low states in three-level $\Lambda$- system at interaction of such systems with laser fields [1]. It is known that the appearance of such dark state has been induced by induction of low-frequency coherence between low states of the atomic system. This fact allows to bring such name as «coherent medium». The changing of characteristics of the optical dense medium and its influence on propagation of multi-frequency laser irradiation was called as effect of electromagnetically-induced transparency (EIT) [2]. In the simple case the EIT takes place when a two-frequency radiation passes through a three-levels atom medium ($\Lambda$-system) and each spectral component interacts with one of two long-lived states of atom with the excited state [3, 4]. Hence the absorption of the optical radiation by the medium becomes negligible when the difference of frequencies of spectral components is equal to the difference of frequencies of the transition between low states in the atom. In other words the dark resonance (or EIT resonance) takes place for these conditions. At the same time the scale of the frequency area of enlightenment (width of the dark resonance) can be equal to several thousandths of a percent of the natural width of the atomic transition line.

It is emphasized that these resonance has been already used successfully for the different applications: the creation of the compact frequency standards or high precisione quantum magnetometers. Moreover the dark resonances are of particular interest for the development of devices for recording and storing quantum information [5] and also quantum logic switches [6]. It is noted that the width of using resonance is directly connected to the reliability of information processing. So the methods of recording and reading of qubits with a high degree of reliability were realized on the basis of the EIT resonance in the atoms inside an optical lattice.

It is obviously that the further progress in the creation of the element basis for quantum computing is connected to realization of the EIT effect in solid-state structure, namely in semiconductor nanostructures – quantum wells and quantum dots. The possibility of creation the structure which has the determined positions of levels of size quantization is important advantage because it allows to adapt the structure almost for any laser generators. The other advantage is fact that in quantum wells have high dipole matrix elements. Due to this different interference effects for condensed state (including related with the EIT resonances) take place.

Therefore the dark resonances are investigated both theoretically [7] and experimentally in semiconductor quantum wells based on InGaAs/AlInAs [8] and GaAs [9] in spite of problems due to broadening of lines of optical transitions. Appearance of the EIT effect due to the coherent state between heavy and light holes in valence band was experimentally investigated in [10]. Existence of the CPT resonance was shown when the system with the three dipole-allowed transitions in asymmetric quantum well was irradiated by two infra-red (IR) fields [11]. Amplification in the emission intensity of 2s state of excitons by generation of the dark states was studied [12]. Appearance of the EIT resonance for interaction of IR radiation with quantum wells was investigated in [13-14]. The dependence of the existence of the EIT resonance on the phases of influencing fields was studied for the four-level inverted Y configuration which was realized in the system of asymmetric quantum well *n*-InGaAs (with the AlGaAs barriers) when it was interacting with three-frequency field [15]. Appearance of the EIT resonance for femtosecond excitation in the structures with array of quantum wells was observed [16]. The observation of EIT in the photonic crystals was realized [17].

The structures consisting of double tunneling-coupled quantum wells also is interesting. The theory of the EIT in the asymmetric double quantum wells was invented [18]. The nonlinear response of light in the presence of a strong field for EIT resonance was theoretically investigated in such structures [19]. However there is some problems in ones. For example it is realization of the dark resonances in the closed scheme of the excitation contour where a corellation between phases of exciting fields is the main condition of existence of the dark resonances.

This work is devoted to research of the dark resonances in semiconductor double tunneling-coupled quantum wells. The possibility to control the dark resonances by coupling field between the energy levels of holes in the valence band is considered in this work. As a result the closed contour of excitation takes place and it gives large potential to control formation of the dark resonances by means of variation the common phase of the contour as well as intensity of coupling field [15, 20-23]. These researches are actual due to possibility of building the devices for record and processing of quantum information in the basis of the dark resonances in semiconductor nanostructures.

**2. Basic equations**

We consider the double tunneling-coupled quantum wells structure, see Fig.1. States $|3\rangle$ and $|4\rangle$ are the combinations of wave functions in each of quantum wells (the barrier between the wells is permeable). The value $2\Delta$ of high levels tunnel splitting depends on probability of tunneling between these wells. Wave functions of the states $|1\rangle$ and $|2\rangle$ overlap each other. It allows to couple the low levels by additional infrared electromagnetic field.

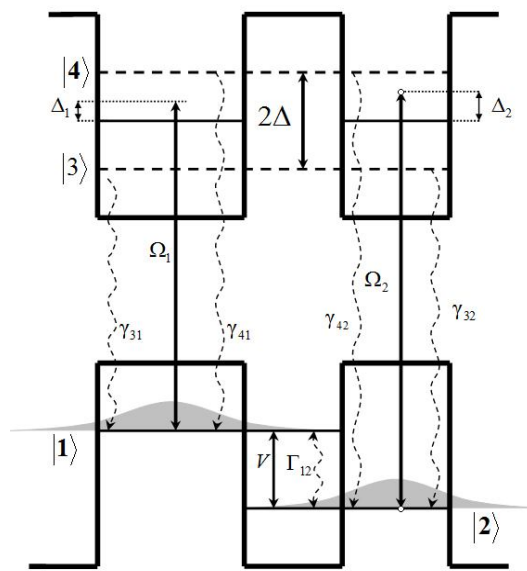

**Fig.1.** The scheme of the energy levels in the double tunneling-coupled quantum wells: $2\Delta$ is tunnel splitting; $\Omega_1$, $\Omega_2$ are Rabi frequencies of optical fields interacting on the transitions $|1\rangle \to |3\rangle(|4\rangle)$ and $|2\rangle \to |3\rangle(|4\rangle)$; $\Delta_1$ and $\Delta_2$ are one-photon detuning of frequencies of optical fields and V is Rabi frequency of infrared field interacting on the transition $|1\rangle \to |2\rangle$.

The set of equations for the density matrix elements $\rho_{ij}$ describing the interaction of this system with three-frequency laser field can be inscribed as

$$\frac{\partial \rho_{ik}}{\partial t} = -\frac{i}{\hbar} \sum_l [H_{il}\rho_{lk} - \rho_{il}H_{lk}] + \sum_{l,m} \Gamma_{ik,lm} \rho_{lm}, \qquad (1)$$

where $H$ is the Hamiltonian, and $\Gamma$ is the relaxation matrix. Hamiltonian $H$ can be represented as $H = H_0 + H_{int}$, where $H_0$ is the Hamiltonian in the absence of laser field

$$H_0 = \sum_{i=1}^{4} E_i |i\rangle\langle i|, \qquad (2)$$

and $H_{int}$ describes the interaction of quantum system with laser field. In resonance approximation

$$H_{int} = \hbar q \Omega_1 e^{-i(\nu_1 t + \varphi_1)} |3\rangle\langle 1| + \hbar \Omega_1 e^{-i(\nu_1 t + \varphi_1)} |4\rangle\langle 1| + \hbar \kappa \Omega_2 e^{-i(\nu_2 t + \varphi_2)} |3\rangle\langle 2| +$$
$$+ \hbar \Omega_2 e^{-i(\nu_2 t + \varphi_2)} |4\rangle\langle 2| + \hbar V e^{-i(\nu_3 t + \varphi_3)} |2\rangle\langle 1| + h.c. \qquad (3)$$

Here $\Omega_1 = \mu_{13} E_1 / 2\hbar$, $q\Omega_1 = \mu_{14} E_1 / 2\hbar$, $\Omega_2 = \mu_{23} E_2 / 2\hbar$ and $\kappa \Omega_2 = \mu_{24} E_2 / 2\hbar$ are the Rabi frequencies, $E_i$ and $\varphi_i$ are the amplitude and phase of $i$-th laser field component with frequency $\nu_i$ ($i = 1, 2$); $\mu_{13}$, $\mu_{14}$, $\mu_{23}$ and $\mu_{24}$ are the dipole moments of transitions $|1\rangle \to |3\rangle$, $|1\rangle \to |4\rangle$, $|2\rangle \to |3\rangle$ and $|2\rangle \to |4\rangle$ respectively; $q = \mu_{14}/\mu_{13}$, $\kappa = \mu_{24}/\mu_{23}$. Also we suppose that infrared laser field with Rabi frequency $V$, frequency $\nu_3$, and phase $\varphi_3(t)$ interacts with the transition $|1\rangle \to |2\rangle$ which has dipole moment $\mu_{12}$.

The relaxation matrix $\Gamma$ are the following: $\Gamma_{22,11} = \Gamma_{11,22} = \Gamma_{22,22} = \Gamma_{11,11} = \gamma_{21} = 2.5 \cdot 10^{-5} \gamma$, $\Gamma_{11,33} = \Gamma_{22,33} = \gamma_{31} = \gamma_{32} = 0.8\gamma$, $\Gamma_{33,33} = \gamma_{31} + \gamma_{32}$, $\Gamma_{11,44} = \Gamma_{22,44} = \gamma_{41} = \gamma_{42} = 0.75\gamma$, $\Gamma_{44,44} = \gamma_{41} + \gamma_{42}$, $\Gamma_{12,12} = \Gamma_{21,21} = \Gamma_{12} = 4\gamma_{21}$, $\Gamma_{13,13} = \Gamma_{31,31} = \Gamma_{13} = 1.92\gamma$, $\Gamma_{14,14} = \Gamma_{41,41} = \Gamma_{14} = 1.8\gamma$, $\Gamma_{23,23} = \Gamma_{32,32} = \Gamma_{23} = 1.92\gamma$, $\Gamma_{24,24} = \Gamma_{42,42} = \Gamma_{24} = 1.8\gamma$, $\Gamma_{43,43} = \Gamma_{34,34} = \Gamma_{34} = 3.41\gamma$ ($\gamma = 1.519 \cdot 10^{11} \sec^{-1} (= 1 meV)$) [24]. In the rotating frame the set of equations for the density matrix is:

$$\dot{\rho}_{11} = iV(\rho_{12} - \rho_{21}) + iq\Omega_1(\rho_{13} - \rho_{31}) + i\Omega_1(\rho_{14} - \rho_{41}) + 2\gamma_{21}(\rho_{22} - \rho_{11}) + 2\gamma_{31}\rho_{33} + 2\gamma_{41}\rho_{44}, \quad (4)$$

$$\dot{\rho}_{22} = iV(\rho_{21} - \rho_{12}) + i\kappa\Omega_2(\rho_{23} - \rho_{32}) + i\Omega_2(\rho_{24} - \rho_{42}) + 2\gamma_{21}(\rho_{11} - \rho_{22}) + 2\gamma_{32}\rho_{33} + 2\gamma_{42}\rho_{44}, \quad (5)$$

$$\dot{\rho}_{33} = iq\Omega_1(\rho_{31} - \rho_{13}) + i\kappa\Omega_2(\rho_{32} - \rho_{23}) - 2(\gamma_{31} + \gamma_{32})\rho_{33}, \quad (6)$$

$$\dot{\rho}_{44} = i\Omega_1(\rho_{41} - \rho_{14}) + i\Omega_2(\rho_{42} - \rho_{24}) - 2(\gamma_{41} + \gamma_{42})\rho_{44}, \quad (7)$$

$$\dot{\rho}_{12} = \left[i(\Delta_2 - \Delta_1 + \dot{\varphi}_3(t)) - \Gamma_{12}\right]\rho_{12} + i\kappa\Omega_2 e^{i\Phi(t)}\rho_{13} + i\Omega_2 e^{i\Phi(t)}\rho_{14} - iq\Omega_1 e^{i\Phi(t)}\rho_{32} - i\Omega_1 e^{i\Phi(t)}\rho_{42} + iV(\rho_{11} - \rho_{22}), \quad (8)$$

$$\dot{\rho}_{13} = \left[i(-\Delta_1 - \Delta + \dot{\varphi}_1(t)) - \Gamma_{13}\right]\rho_{13} + i\kappa\Omega_2 e^{-i\Phi(t)}\rho_{12} - iVe^{-i\Phi(t)}\rho_{23} - i\Omega_1\rho_{43} + iq\Omega_1(\rho_{11} - \rho_{33}), \quad (9)$$

$$\dot{\rho}_{14} = \left[i(-\Delta_1 + \Delta + \dot{\varphi}_1(t)) - \Gamma_{14}\right]\rho_{14} + i\Omega_2 e^{-i\Phi(t)}\rho_{12} - iVe^{-i\Phi(t)}\rho_{24} - iq\Omega_1\rho_{34} + i\Omega_1(\rho_{11} - \rho_{44}), \quad (10)$$

$$\dot{\rho}_{23} = \left[i(-\Delta_2 - \Delta + \dot{\varphi}_2(t)) - \Gamma_{23}\right]\rho_{23} + iq\Omega_1 e^{i\Phi(t)}\rho_{21} - iVe^{i\Phi(t)}\rho_{13} - i\Omega_2\rho_{43} + i\kappa\Omega_2(\rho_{22} - \rho_{33}), \quad (11)$$

$$\dot{\rho}_{24} = \left[i(-\Delta_2 + \Delta + \dot{\varphi}_2(t)) - \Gamma_{24}\right]\rho_{24} + i\Omega_1 e^{i\Phi(t)}\rho_{21} - iVe^{i\Phi(t)}\rho_{14} - i\kappa\Omega_2\rho_{34} + i\Omega_2(\rho_{22} - \rho_{44}), \quad (12)$$

$$\dot{\rho}_{34} = \left[2i\Delta - \Gamma_{34}\right]\rho_{34} - iq\Omega_1\rho_{14} - i\kappa\Omega_2\rho_{24} + i\Omega_1\rho_{31} + i\kappa\Omega_2\rho_{32}, \quad (13)$$

Here $\Delta_1 = \nu_1 - (\omega_{31} + \omega_{41})/2$ and $\Delta_2 = \nu_2 - (\omega_{32} + \omega_{42})/2$ are one-photon detuning of laser frequencies from transitions $|1\rangle \to |3\rangle(|4\rangle)$ and $|2\rangle \to |3\rangle(|4\rangle)$, $\omega_{ij}$ is the frequency of transition between levels $i$ and $j$, $\Phi(t) = \varphi_1(t) - \varphi_2(t) - \varphi_3(t)$ is the total phase.

## 3. Results and discussion

Equations (4) - (13) describe dynamics of atomic populations $\rho_{11}, \rho_{22}, \rho_{33}, \rho_{44}$ as well as the optical coherences $\rho_{13}, \rho_{14}, \rho_{23}, \rho_{24}$ in dependence of the Rabi frequencies $\Omega_1$ and $\Omega_2$ for optical fields, the Rabi frequency V of coupling field and the phase $\Phi$. Further we will consider the stationary solution where all time derivatives are equal to zero:

$$\frac{\partial \rho_{ij}(t)}{\partial t} = 0 \ (i, j = 1 \div 4), \ \frac{\partial \varphi_n(t)}{\partial t} = 0 \ (n = 1 \div 3). \quad (14)$$

Then the set of differential equations (4) - (13) transforms to the set of ordinary linear equations.

First of all we note that the presence of closed contour of excitation radically changes characteristics of the medium [20]. Closing of atomic contour can be realized by radiofrequency field, as in the case of three-level $\Lambda$-system [21], or optical fields, as in the case of double $\Lambda$-system [22]. In our case closing of the contour of excitation is realized through two channels: tunneling between the excited states in both quantum wells, and by the infrared field acting on the transition $|1\rangle \leftrightarrow |2\rangle$, see Fig.1.

Total phase $\Phi = \varphi_1 - \varphi_2 - \varphi_3$ is a parameter characterizing the closed excitation contour in the considered structure. Populations and coherences are crucially depending on this phase.

The dependencies of populations $\rho_{11} + \rho_{22}, \rho_{33}, \rho_{44}$ on two-photon detuning $\delta = (\Delta_1 - \Delta_2)/2$ for different values $\Phi$ are presented on fig.2. We can see that the CPT resonance take place in the double tunneling-coupled quantum wells for $\Phi = 0$. Actually almost all population is concentrated on the low levels (Fig.2 (a) – the solid curve) in the case of two-photon resonance ($\delta = 0$). At the same time there is a narrow resonance for the population of high levels. It is a specific well on population curves of the excited states $|3\rangle, |4\rangle$ for close to zero values of two-photon detuning ($\delta \approx 0$) (see Fig.2 b, c). It is possible to control the contrast of dark resonances by means of variation of the phase $\Phi$ like in the case of atomic systems [20-22]. So the amplitude of dark resonance decreases for $\Phi = \pi/4$ (dot curve) and disappears for $\Phi = \pi/2$ (dash-dot curve) which means the destruction of the dark state. Physically such destruction of coherent population trapping is a result of absence of dark superposition states of the low levels for nonzero values of the phase over the excitation contour.

Also the difference of populations $\rho_{33}$ and $\rho_{44}$ on the Fig.2(b) and 2(c) is a result of difference in Rabi frequencies of the transitions $|1\rangle \rightarrow |3\rangle(|4\rangle)$ and $|2\rangle \rightarrow |3\rangle(|4\rangle)$ due to different dipole moments of appropriate transitions (q=k=0.8).

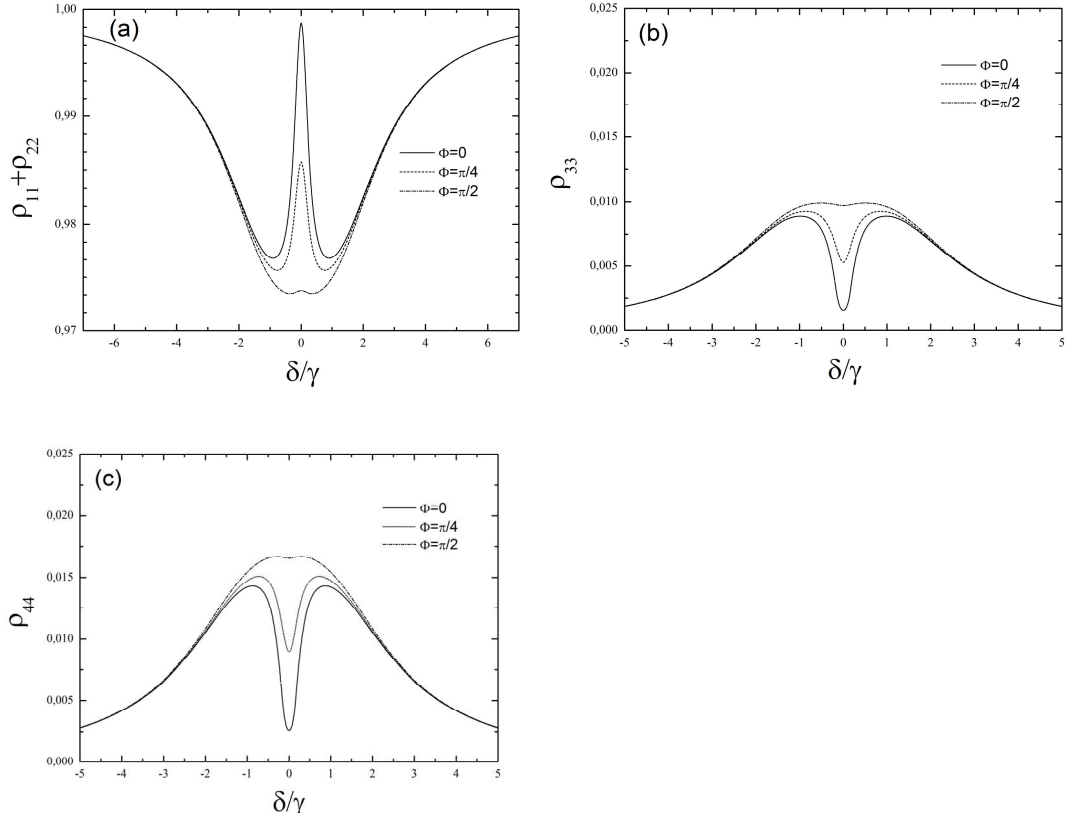

**Fig.2.** The dependence of populations $\rho_{11}+\rho_{22}$ - (**a**), $\rho_{33}$ - (**b**), $\rho_{44}$ - (**c**) on two-photon detuning for three values of summary phase $\Phi$. Here $\Omega_1 = \Omega_2 = V = 0.25\gamma$, $\Delta = \gamma$, $q = \kappa = 0.8$.

Let us consider the influence of the Rabi frequency $V$ of the coupling infrared field on the existence of the dark resonance in the system of tunneling-coupled quantum wells. Choose the value of $\Phi = \pi/2$. If $V \simeq \Omega_1, \Omega_2$, then there is no CPT resonance (the dash-dot curve on the Fig.3,4). When Rabi frequency $V$ decreases, the contrast of the dark resonance is restored (the dot and dash curves on the Fig.3,4). This is due to destruction of the closed contour of excitation $|1\rangle \to |3\rangle(|4\rangle) \leftrightarrow |2\rangle \leftrightarrow |1\rangle$ for close to zero Rabi frequencies of infrared radiation ($V \to 0$). As a result, the scheme of excitation on the Fig.1 for $V \to 0$ transforms similarity of $\Lambda$-system where the existence of the dark resonance does not depend on the phase $\Phi$.

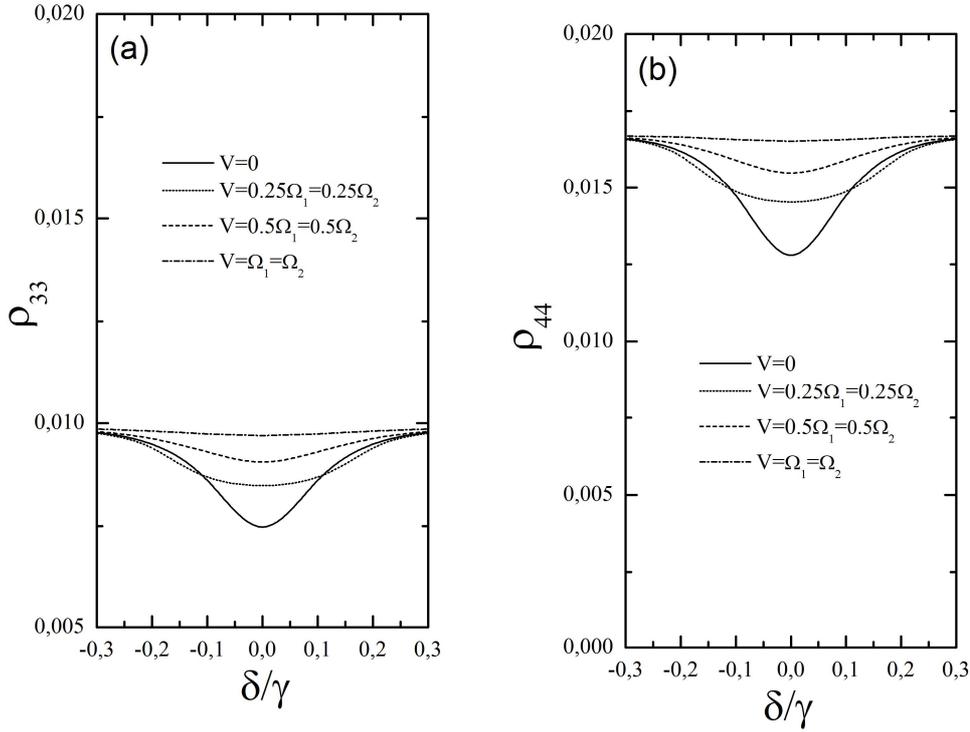

**Fig.3.** Dependence of populations $\rho_{33}$ (**a**) and $\rho_{44}$ (**b**) on two-photon detuning for different values of Rabi frequency of coupled field. Here $\Omega_1 = \Omega_2 = 0.25\gamma$, $\Delta = \gamma$, $q = \kappa = 0.8$, $\Phi = \dfrac{\pi}{2}$.

Further we are considering dispersion and absorption of probe field for the transitions $|1\rangle \to |3\rangle (|4\rangle)$ in the system of double tunneling-coupled quantum wells (Fig.1). We suppose that strong field of optical radiation interacts on the transitions $|2\rangle \leftrightarrow |3\rangle$, $|2\rangle \leftrightarrow |4\rangle$. Since the imaginary part of optical coherence $\mathrm{Im}(\rho_{13} + \rho_{14})$ is proportional to absorption, and the real part $\mathrm{Re}(\rho_{13} + \rho_{14})$ determines the index of refraction (or dispersion), therefore we must solve the set of equations (4) - (13) for the stationary case to find the non-diagonal elements of density matrix $\rho_{13}$, $\rho_{14}$ for the condition $\Omega_1 \ll \Omega_2$.

There are the dependencies of absorption (Fig.4a) and dispersion (Fig.4b) of probe field on the transition $|1\rangle \to |3\rangle (|4\rangle)$ for tree values of the phase $\Phi$. There is qualitative agreement of the dependencies of absorption and index of refraction under CPT conditions (the solid curve on the Fig.4) with the case of probe field in tree-level $\Lambda$-system [25]

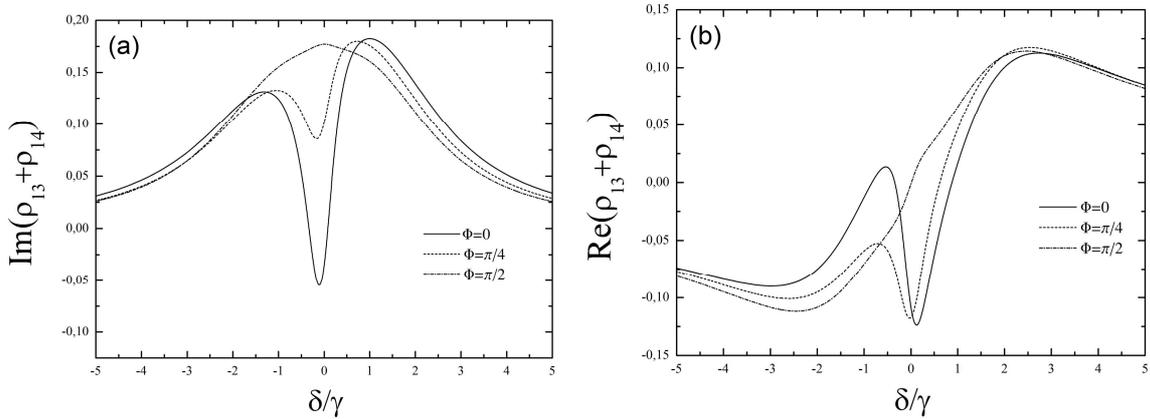

**Fig.4.** Dependence of $\text{Im}(\rho_{13}+\rho_{14})$ **(a)** (absorption coefficient of the probe field) and $\text{Re}(\rho_{13}+\rho_{14})$ **(b)** (dispersion coefficient of the probe field) on two-photon detuning for three values of phase $\Phi$. Here we took $\Omega_1 = V = 0.25\gamma$, $\Omega_2 = 3\Omega_1$, $\Delta = \gamma$, $q = \kappa = 1$.

On the other hand on the contrary with simple $\Lambda$-system, the dispersion coefficient of the probe field reaches a maximum near the value $\delta \simeq 0$, and absorption gives way to amplification. At the same time in simple $\Lambda$-system the maximum of dispersion is observed for nonzero absorption and $|\delta| \neq 0$. In other words the maximum of dispersion is almost coinciding with the maximum of amplification of the probe field by the medium in the case of tunneling-coupled quantum wells. The last occurs is due to presence of closed excitation contour in the system. It allows to redistribute the radiation power and to amplify the probe field for maximum value of dispersion coefficient.

Some asymmetry in the dependencies of absorption and dispersion coefficients on two-photon detuning is due to inequality of the matrix elements of dipole interaction operator for the transitions $|1\rangle \to |3\rangle$ and $|1\rangle \to |4\rangle$. Significant dispersion (almost like for zero phase) is observed for the value of phase $\Phi = \pi/4$ (the dot curve) in the area $\delta \simeq 0$, but the probe field absorption increases abruptly. This is the consequence of CPT destruction. Finally, when the value of phase $\Phi = \pi/2$ and CPT is destructed, the dependencies of absorption and dispersion spectrums of probe field correspond to two-level atom model (the dash-dot curves).

On the fig. 5 the dependence of dispersion on a value of two-photon detuning for different values of coupling field and phase $\Phi = \pi/4$ is presented. From this figure one can see that resonant character of this specific decrease when the intensity of coupling field increases.

So CPT takes place actually for $\Phi = \pi/4$ when value of coupling field $V$ is $V = 0$ (solid curve) in the system of two quantum wells. Accordingly the sharp dispersive specific takes place in the aria $\delta \simeq 0$ for dependence of dispersion coefficient. Amplitude of the dispersive singularity decreases a bit when the Rabi frequency of coupling field increases. This is due to CPT destruction for nonzero phase of the excitation contour (the dot, dash and dash-dot curves). Any specifics of dispersion coefficient disappear for the values of Rabi frequency of coupling field $V \gg 3\Omega_1$ (the dot curve).

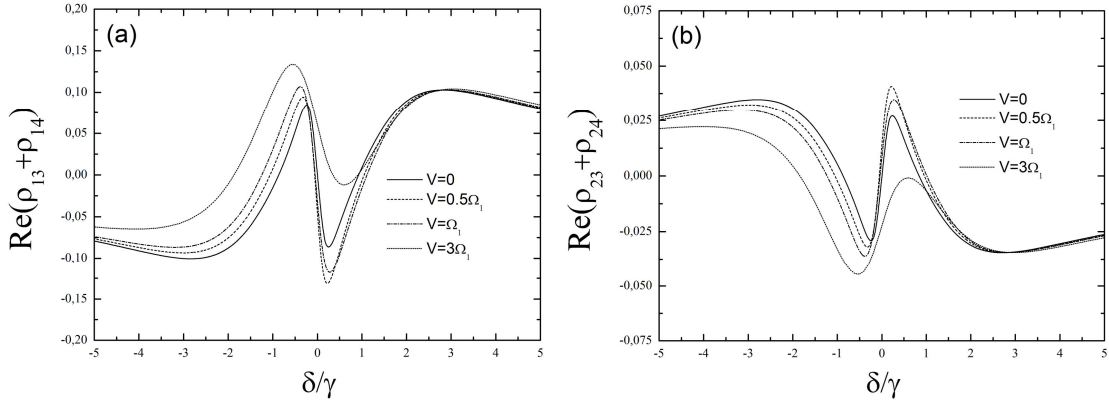

**Fig.5.** The dependence of dispersion of the probe field $\mathrm{Re}(\rho_{13} + \rho_{14})$ (**a**) and of the strong field $\mathrm{Re}(\rho_{23} + \rho_{24})$ (**b**) on two-photon detuning for different values of Rabi frequency of coupling field. Here $\Omega_1 = 0.25\gamma$, $\Omega_2 = 3\Omega_1$, $\Delta = \gamma$, $q = \kappa = 1$, $\Phi = \dfrac{\pi}{4}$.

On the Fig.6 the dependence of summary population $\rho_{33} + \rho_{44}$ on two-photon detuning for the different values of splitting $\Delta$ which characterize thickness of the potential barrier is presented. We chose two values of phase of the contour of excitation: $\Phi = 0$ (Fig.6(a)) when CPT takes place and $\Phi = \pi/2$ (Fig.6(b)) when the dark resonance is destructed. From the fig.6 one can see that amplitude of the dark resonance (specific well on population curves of the excited states $|3\rangle, |4\rangle$ for close to zero values of two-photon detuning) decreases when $\Delta$ is increasing for $\Phi = 0$. Such amplitude decreasing is due to the increasing of one-photon detunings $\Delta_1$, $\Delta_1$ together with splitting $\Delta$. This leads to weakening of interaction of laser radiation and the two quantum wells system (Fig.1) on the transitions $|1\rangle \to |3\rangle (|4\rangle)$ and $|2\rangle \to |3\rangle (|4\rangle)$.

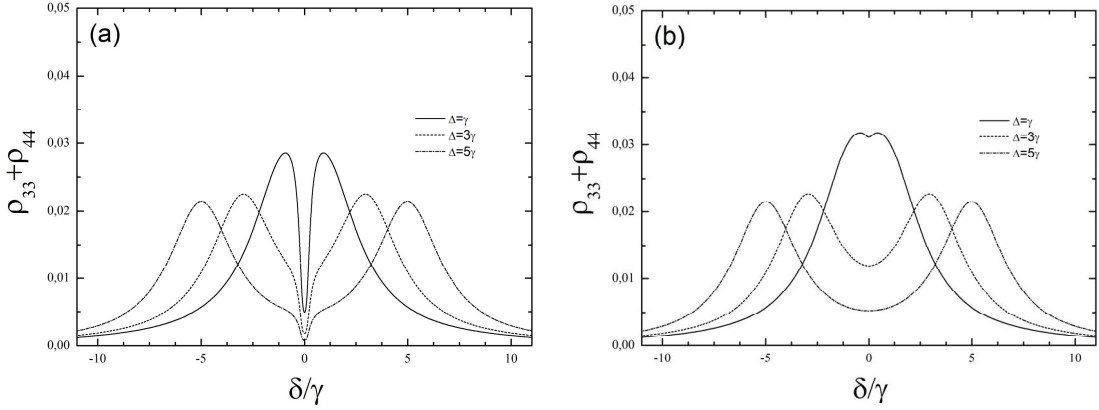

**Fig.6.** The dependence of summary population $\rho_{33}+\rho_{44}$ of excited states on splitting $\Delta$ for $\Phi=0$ (**a**), $\Phi=\dfrac{\pi}{2}$ (**b**). Here $\Omega_1=\Omega_2=V=0.25\gamma$, $q=\kappa=1$.

The dependence of populations on the parameter $q$ is presented on the Fig.7. Let us call the superpositions of states $|1\rangle$ and $|2\rangle$ what are not connected with states $|3\rangle$ and $|4\rangle$ by inducted transitions as $|\Psi_{NC}^3\rangle$ and $|\Psi_{NC}^4\rangle$ respectively. These wave functions are the same in the case of two-photon resonance and for $q=1$. So both CPT excitation channels interfere constructively. If q=-1, the states $|\Psi_{NC}^3\rangle$ and $|\Psi_{NC}^4\rangle$ are orthogonal. This fact leads to vanish of the dark resonance.

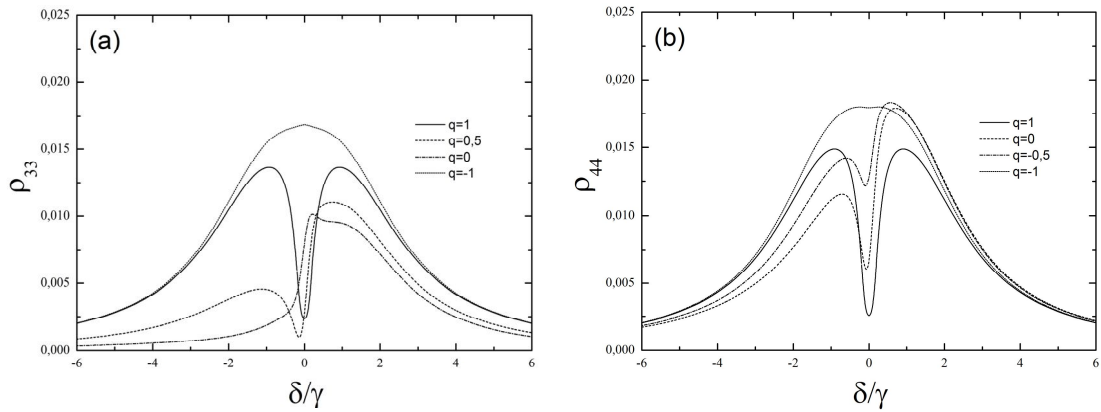

**Fig.7** The form of the dark resonance for populations according to the q parameter: (**a**) - $\rho_{33}$, (**b**) - $\rho_{44}$. Here $\Omega_1=\Omega_2=V=0.25\gamma$, $\Delta=\gamma$, $\kappa=1$, $\Phi=0$.

The dependence of summary population of the levels $|3\rangle$ and $|4\rangle$ on the Rabi frequency $V$ of coupling IR-field is presented on Fig.8. We can see that the width of dark resonance is determined by Rabi frequencies of optical fields and by decay of low-frequency coherency, see Fig.8(a). At the same time the form of the dark resonance varies radically when coupling field is switched on (Fig.8(b)) and depends on the ratio of Rabi frequencies of optical fields. This asymmetry is due to the pumping of power by means of coupling field.

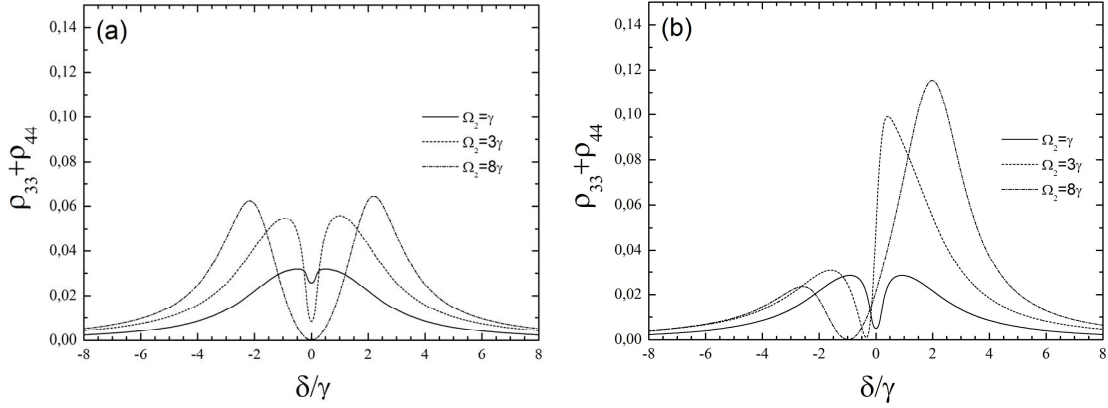

**Fig.8.** The form of the dark resonance for summary population $\rho_{33}+\rho_{44}$ according to Rabi frequency $V$ of coupling field: **(a)** - $V=0$, **(b)** - $V=0.25\gamma$. Here $\Omega_1=0.25\gamma$, $\Delta=\gamma$, $q=\kappa=1$, $\Phi=0$.

### 4. Conclusions

We considered the interaction of three laser fields (two optical fields and one infra-red coupling field) forming the closed excitation contour in double tunneling-coupled quantum wells. It is established that the dark resonance can arise and vanish depending on the total phase $\Phi$ over the excitation contour. For $\Phi=0$ the amplitude of the dark state is maximal and for $\Phi=\pi/2$ the dark resonance is absence. We found that the dispersion coefficient laser fields of optical range in the area of dark resonance at $\Phi=\pi/4$ have different resonance peculiarities which vanish when the amplitude of coupled field increase. It is worth to note that the absorption of the laser radiation passing through this structure was analyzed depending on the two-photon detuning. However when lasers have determined wavelengths, we can realize equivalent effect toward the variation of laser wavelength by means of the impact of constant electric field and variation of it value which leads to shift of the levels of size quantization. In conclusion it worth to say that performed research is actual to the development of quantum information recording and processing.

This work was supported by Federal Special-purpose Program "Science and Science-pedagogical personnel of Innovation Russia on 2009-2013", by Russian President Grant for Young Candidates of Sciences (project MK-5318.2010.2), by Russian Foundation for Basic Researches.

## 5. References

[1] B.D. Agapiev, M.B. Gorniy, B.G. Matisov and Yu.V. Rozhdestvensky // UFN **163**, 1 (1993)

[2] S. Harris // Physics Today **50**, 1997 (36); M. Fleischhauer, A. Imamoglu, J. P. Marangos // Rev.Mod.Phys.**77**, 633 (2005)

[3] M.B. Gorny, B.G. Matisov and Yu.V. Rozhdestvenski // ЖЭТФ **95,** 81 (1989).

[4] O.A. Kocharovskaya, Ya. I. Hanin // JETP **90**, 1610 (1986).

[5] A. Nazarkin, R. Netz, and R. Sauerbrey // Phys. Rev. Lett. **92**, 043002 (2004)

[6] J. H. Shapiro and F. N. C. Wong // Phys. Rev. A **73**, 012315 (2006)

[7] M. Lindberg, R. Binder // Phys.Rev.Lett. **75**, 1403 (1995);

[8] G.B. Serapiglia, E. Paspalakis, C. Sirtori, et. al. // Phys.Rev.Lett. **84**, 1022 (2000);

[9] M.C. Phillips, H. Wang, I. Rumyantsev. et. al // Phys.Rev.Lett. **91**, 183602 (2003).

[10] M.C. Phillips, H. Wang // Optics Letters **28**, 831 (2003).

[11] J.F. Dynes, M.D. Frogley, J. Rodger, et. al. // Phys. Rev. B **72**, 085323 (2005).

[12] S.M. Sadeghi, W.Li // Phys. Rev. B **72**, 075347 (2005);

[13]S.M. Sadeghi, W. Li, X.Li, et. al., // Phys.Rev B **74**, 161304 (2006);

[14] Z. Dutton, M. Murali, W.D. Oliver, et. al., // Phys.Rev B **73**, 104516 (2006).

[15] A. Joshi // Phys. Rev. B **79**, 115315 (2009).

[16] S.M. Ma, H. Xu, B.S. Ham // Optics Express **17**, 148902 (2009)

[17] X. Yang, M. Yu, D.L. Kwong, et. al. // Phys.Rev.Lett. **102**, 173902 (2009).

[18] L.Silvestri, F. Bassani, G. Czajkowski, et. al // Eur. Phys. J. B. **27**, 89 (2002)

[19] L.Silvestri and G. Czajkowski // Phys. Stat. Sol. C **5**, 2412 (2008)

[20] D.V. Kosachiov, B.G. Matisov, Yu.V. Rozhdestvensky // Opt. Commun. **85**, 209 (1991).

[21] D.V. Kosachiov, B.G. Matisov, Yu.V. Rozhdestvensky // J. Phys. B. **25**, 2473 (1992).

[22] D.V. Kosachiov, B.G. Matisov, Yu.V. Rozhdestvensky // Europhys. Letters. **22**, 11 (1993).

[23] S. Kajari-Schored, G. Morigi, S. Franke-Arnold, et.al // Phys.Rev A **75**, 013816 (2007).

[24] W.-X. Yang, A.-X. Chen, T.-T. Zha and R.-K. Lee // J. Phys. B: **42** 225501 (2009)

[25] M.O. Scully, M.S. Zubayri, M.: Fizmatkniga, 2003, 512 c.